# Magnetic wire active microrheology of human respiratory mucus

**Milad Radiom[1,2], Romain Hénault[1], Salma Mani[3], Aline Grein Iankovski,[1] Xavier Norel[3], Jean-François Berret[1]\***

[1]*Université de Paris, CNRS, Matière et systèmes complexes, 75013 Paris, France*
[2]*Institute for Food, Nutrition and Health, D-HEST, ETH Zürich, Zürich, Switzerland*
[3]*Université de Paris, INSERM, UMR-S 1148, CHU X. Bichat, 75018 Paris, France; Paris 13 University, 93430 Villetaneuse, France*

**Abstract** Mucus is a viscoelastic gel secreted by the pulmonary epithelium in the tracheobronchial region of the lungs. The coordinated beating of cilia moves mucus upwards towards pharynx, removing inhaled pathogens and particles from the airways. The efficacy of this clearance mechanism depends primarily on the rheological properties of mucus. Here we use magnetic wire based microrheology to study the viscoelastic properties of human mucus collected from human bronchus tubes. The response of wires between 5 and 80 µm in length to a rotating magnetic field is monitored by optical time-lapse microscopy and analyzed using constitutive equations of rheology, including those of Maxwell and Kelvin-Voigt. The static shear viscosity and elastic modulus can be inferred from low frequency ($3\times10^{-3}$ – 30 rad s$^{-1}$) measurements, leading to the evaluation of the mucin network relaxation time. This relaxation time is found to be widely distributed, from one to several hundred seconds. Mucus is identified as a viscoelastic liquid with an elastic modulus of 2.5 ± 0.5 Pa and a static viscosity of 100 ± 40 Pa s. Our work shows that beyond the established spatial variations in rheological properties due to microcavities, mucus exhibits secondary inhomogeneities associated with the relaxation time of the mucin network that may be important for its flow properties.

## 1 - Introduction

The mammalian respiratory system is equipped with several defense mechanisms along the airways which serve to reduce the adverse effects of inhaled particulate matter. The first of these mechanisms is a physical barrier known as mucociliary clearance (MCC) in the tracheobronchial region.[1,2] MCC consists of a recurring gel layer (mucus) secreted by the goblet cells and submucosal glands, and of metachronal waves of the cilia which expel the mucus from the airways. Mucus barrier has a thickness of about 2 – 5 µm, and is mainly composed of water and a network of mucin glycoproteins (5 wt. %).[1] Much of the protective role of mucus is due to its adhesive properties as a result of high sugar concentration and a mesh structure.[3,4] The efficacy of MCC depends primarily on the rheological properties of mucus.[5] In this regard, several of pulmonary diseases such as asthma, cystic fibrosis and chronic obstructive pulmonary disease (COPD) are known to increase the mucus viscosity with secondary effects such as accumulation of pathogens and other inhaled particles leading to inflammation and epithelium damage.[6-8] Besides it was shown recently that the elastic modulus of expectorated samples from patients with asthma, COPD and cystic fibrosis, when measured at large deformation increases in this order compared to healthy mucus, suggesting that rheology could be used as a biomarker for chronic bronchial disease.[9]

Rheological measurements of human pulmonary mucus are however difficult to carry out. One reason is the minute quantity of sample available from healthy individuals or diseased patients, typically of the order of a few microliters. Another reason is related to the mucus hydrogel structure, which contains nano-[10,11] and microcavities[12,13] in the range of 100 nm to 5 µm. These characteristics limit the application of both macro- and microrheology techniques to study the rheological properties of mucus.[1,5] Motivated by the search for novel drug carriers for pulmonary therapies, bead tracking experiments, actually similar to those realized in passive microrheology, have been performed extensively in the recent years.[10-12,14-23] These studies have demonstrated significant dependencies of the bead thermal diffusion on particle size and coating. It is now established that nanoparticles coated with poly(ethylene glycol) polymers can diffuse rather freely throughout the mucus microcavities. To overcome the difficulty related to macrorheology volume requirement, researchers have focused on sputum samples.[7,9,11,13,14,18,24-31] The sputum samples are generally obtained from patients with serious lung conditions, such as cystic fibrosis and COPD. The collection methods are invasive and present a risk of contamination with saliva or blood.[10,26]

We compiled sputum and mucus rheology data coming from 23 surveys.[9-11,13,14,16,24,26-39] Data on cystic fibrosis sputum[7,9,13,14,24,28-31,33] shows a gel-like behavior with the elastic modulus $G'(\omega)$ larger than the loss modulus $G''(\omega)$ over the frequency range 0.01 – 100 rad s$^{-1}$. At the angular frequency $\omega = 1$ rad s$^{-1}$, it is found that the averaged values of $G'$ and $G''$ for cystic fibrosis sputum are 4.4 ± 1.9 Pa and 3.0 ± 2.0 Pa, respectively, whereas its shear viscosity is estimated to be 67 ± 34 Pa s at a shear rate of 1 s$^{-1}$. The mucus data are in contrast less numerous and show viscosity values with larger uncertainties.[10,31,34,38] The substantial extend of variation in the reported values is due to large inter-sample variations. This property of mucus makes interpretation of rheological properties difficult, especially for the design and development of mucolytic agents. From this literature compilation, it was also concluded that both expectorated (sputum) and native mucus have the characteristics of a viscoelastic gel.[40,41]

In this work, we investigate mucus rheological properties at length scales that are larger than the previously investigated nano- and micro-cavities.[42,43] We use magnetic rotational spectroscopy (MRS) which is based on the time-lapse microscopy tracking of magnetically actuated micron-sized wires.[44-48] The choice of MRS, an active microrheology technique over a passive one to study human mucus was motivated by recent studies that we have performed on model viscoelastic liquids and gels.[49] There, we showed that for fluids of viscosity higher than few tens of Pa s, the orientational Brownian fluctuations of wires were small, and did not allow to determine satisfactorily the wire mean-squared angular displacement, the equivalent of the mean-squared displacement for beads. As data from the existing literature revealed in some cases mucus viscosities of several hundred Pa s[32,35,50], it was therefore anticipated that active microrheology would be more suitable here. To this, it should be noted that wire-based active microrheology allows probing material properties at frequencies down to $10^{-3}$ rad s$^{-1}$, and thereby to explore long time relaxation dynamics.

We have investigated human mucus from two collection strategies. *Ex Vivo* mucus samples were either directly collected from human bronchus tube after surgery (we call these samples early *Ex Vivo* mucus) or from the culture of human bronchus tissue (late *Ex Vivo* mucus). The collected volumes in the former were in the milliliter range, while the volumes secreted in the latter case were in the range of 10 – 30 µl, which is sufficient for MRS. Wires of length 5 – 80 µm, *i.e.* longer than the typical mucus microcavities, and diameter 0.7 – 2 µm were added to mucus and submitted to an external magnetic field of increasing frequency. The wire response was monitored by time-lapse optical microscopy and analyzed using constitutive equations of rheology, including Maxwell and Kelvin-Voigt.[44,51,52] In particular, with measurements made at frequencies in the range $3 \times 10^{-3}$ – 30 rad s$^{-1}$, the static shear viscosity and elastic modulus were inferred,[51,52] leading to the evaluation of the mucin network relaxation time. In what follows, we present the results obtained with mucus and discuss the presence of viscoelastic heterogeneities in terms of relaxation times.

## 2 - Materials and Methods

**2.1 - Mucus collection**

Early and late *Ex Vivo* mucus were collected as follows. A human lung was received after transplantation surgery from the Hospital Bichat-Claude-Bernard in Paris (France). In the anatomical pathology department of the same hospital, bronchus tubes of the lung were initially inspected to ensure that they were in a healthy state and that mucus could be collected. Tweezers were used to collect fresh mucus directly from within the healthy bronchus tubes (Fig. 1a and 1b). This mucus is called early *Ex Vivo*. To prepare late *Ex Vivo* mucus, a part of a bronchus was then excised to tubes of length 7.5 – 10 mm and diameter 5 – 10 mm. These tubes were further cut into several smaller rings, as shown in Fig. 1c. The rings were immersed in a Tyrode physiological solution and incubated for 18 h inside a $CO_2$ incubator at 37 °C. During this incubation period, the epithelial mucous (goblet) cells continue to produce mucus. A schematic of the cultured tissue is depicted in Fig. 1d. It is composed of ciliated cells as well as mucus secreting goblet cells. The secreted mucus whose volume was in the range of 10 – 30 µl was then collected using a pipette or tweezers, and kept at 4 °C until microrheology experiments within 24 h after collection which were performed in Université de Paris (**Supporting Information S1**). This procedure was applied to lungs received from 8 patients who underwent operations between January and April 2019. Mucus samples were obtained from the hospital approximately every two weeks, with early and late *Ex Vivo* samples coming each time from a unique patient. Mucus samples were studied by MRS shortly after reception and data on wire rotation as a function of the frequency were acquired. The data were then compared to those of previous patients. As the microrheology results between the batches did not reveal any notable differences, the data of the patients were pooled together and treated equally (**Supporting Information S2**). The mucus collection and bronchial culture procedures described above were approved by the Institutional Review Board (trial registration number IRB 00006477) of INSERM and Assistance Publique - Hôpitaux de Paris and by the Ethics Review Committee for Biomedical Research projects of University Hospital Groups (CEERB of GHU Nord). Prior to surgery, the patients signed a written consent form to participate in scientific studies. These investigations are in accordance with the principles of the Declaration of Helsinki,[53] which sets out ethical principles to guide physicians and other participants in medical research involving human subjects.

### 2.2 - Mucus dry mass

When the mucus samples were in sufficient quantity (> 70 µL) the dry mass was also quantified. For this purpose, the remaining mucus from sample preparation for MRS was heated to 150 °C for 1 h and then weighed to obtain the ratio of the dry mass to the total mass. Experiments were performed in triplicate on the *early Ex Vivo* mucus, leading to mass ratios equal to 4.3 wt. %, 5.7 wt. % and 5.3 wt. % (average 5.1 ± 0.6 wt. %), in good agreement with literature.[1]

### 2.3 - Synthesis of magnetic wires

Superparamagnetic maghemite ($\gamma$-$Fe_2O_3$) nanoparticles with a mean diameter of 13.2 nm and a dispersity of 0.23 were used. The nanoparticles were initially coated with poly(acrylic) acid polymers[54,55] (Aldrich, France) of molecular weight 2100 g mol$^{-1}$ and molar mass dispersity $Đ$ = 1.8.[56] The wires were fabricated using an electrostatic co-assembly process between the negatively charged particles and poly(diallydimethyl ammonium chloride) (PDADMAC), a cationic polymer acting as a glue for the nanoparticles. The degree of polymerization of PDADMAC was determined from the number-average molecular weight $M_n$ = 7600 g mol$^{-1}$ by size exclusion chromatography to be 50, with dispersity of $Đ$ = 3.5[56,57]. The wires were synthesized by desalting 1M $NH_4Cl$ mixed nanoparticle-polymer solutions using a Slide-a-Lyzer® dialysis cassette (Thermo Fisher Scientific, France) with molecular weight cut-off of 10000 g mol$^{-1}$, the dialysis being done at pH 8. The anisotropic shape of the micron-sized wires was induced by applying a magnetic field of 0.3 T during dialysis[54] (**Supporting Information S3**).

### 2.4 - Characterization of magnetic wires

The geometrical characterization of the wires was performed by measuring their length $L$ and diameter $D$ using an optical microscope (Olympus IX73, objective 100×) coupled with a CCD camera (QImaging, EXi Blue) supported by the software Metaview (Universal Imaging). The 100× objective has a numerical aperture of 1.3 and a lateral spatial resolution of 280 nm. From these measurements, it was found that the diameter $D$ increases slightly with the length according to: $D(L) = 0.670L^{0.198}$ (**Supporting Information S4**). This result is attributed to the partial aggregation of the pre-formed parallel wires during synthesis and caused by magnetic dipole interaction between wires. The anisotropy of magnetic susceptibility $\Delta\chi$ characterizing the magnetic response of the wires was determined in independent measurements (as explained in Section 2.6) using water-glycerol mixtures of known viscosity.[44] For these wires $\Delta\chi$ was found to be equal to 2.3 ± 0.5 (**Supporting Information S4&S5**).

## 2.5 - Sample preparation and wire seeding for magnetic rotational spectroscopy experiments

All sample preparations were performed inside a laminar flow hood in a cell culture room. The vials containing the mucus were shut at all time and only opened when wires were added to mucus and prior to deposition on glass coverslip. This step was necessary to minimize the evaporation of the water content of mucus. We adhered a dual adhesive Gene Frame (thickness 250 µm, Abgene, Advanced Biotech) to a cover slip. Afterwards, 0.5 µl of a solution containing magnetic wires was added to 10 – 30 µl of mucus. The sample was vortexed softly and deposited in the center of the Gene Frame. The sample was then covered by a second coverslip and hermetically sealed. An example of wire distribution in mucus is shown in Fig. 1e. We observed that the wires spread somewhat uniformly in the sample. Additionally, wires could be found over the full thickness of mucus sample. MRS measurements were performed on wires located in the center of the sample to reduce wall interaction effects. Occasionally we found large number of wires co-located in one region of mucus. These wires were not selected for microrheology investigation due to possible hydrodynamic or magnetic interactions. Fig. 1e further shows that the mucus sample interacts nonuniformly with passing phase-contrasted light in local regions. This observation manifests that mucus has a heterogeneous material distribution, most likely related to the microcavities mentioned in the introduction.

## 2.6 – Magnetic Rotational Spectroscopy applied to mucus samples

The magnetic wires used in this study have lengths between 5 and 80 µm and diameters between 0.7 and 2 µm. The cell containing the mucus sample with embedded wires was introduced into a homemade device generating a rotational magnetic field from two pairs of coils (resistance 23 Ω) working with a 90°-phase shift. An electronic set-up allowed measurements in the frequency range $\omega = 10^{-3} – 10^2$ rad s$^{-1}$ and at a magnetic field $B = 10.3$ mT (**Supporting Information S6**). The wire rotation was monitored in phase contrast optical microscopy with a 20× objective. For each angular frequency, a movie was recorded for a period of time of at least $10/\omega$ and then treated using the ImageJ software to retrieve the wire rotation angle $\theta(\omega, t)$ and the average rotation velocity $\Omega(\omega) = <d\theta(\omega, t)/dt>_t$. The above experimental procedure was applied to a total of 98 wires, 58 (resp. 40) in early (resp. late) *Ex Vivo* mucus, each of which was studied at 4 – 5 different angular frequencies.

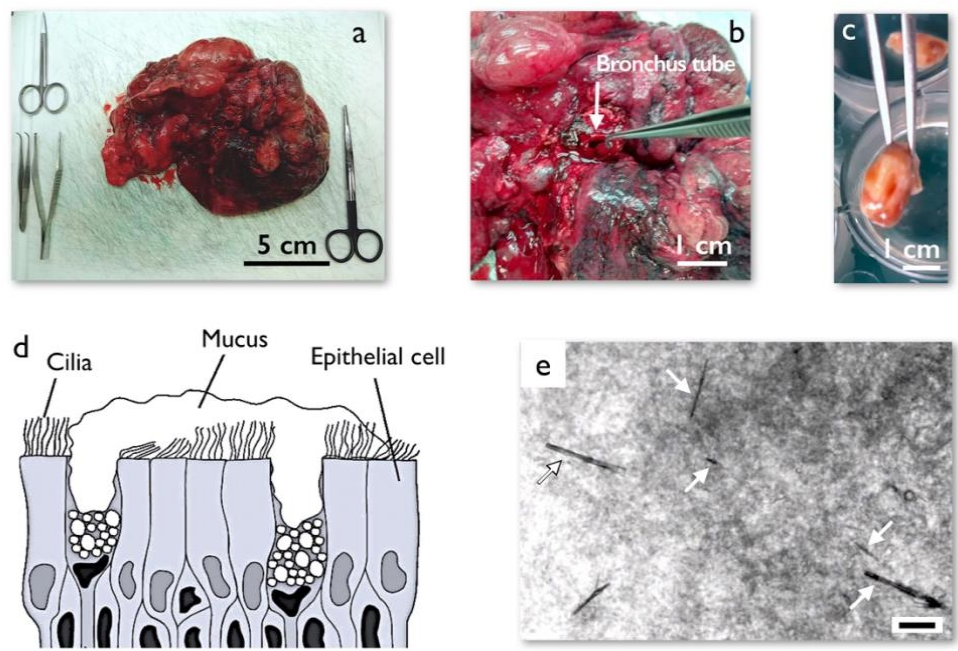

*Figure 1*: a) Piece of a human lung after surgery obtained from the anatomical pathology department at Hospital Bichat-Claude-Bernard, Paris (France). b) Dissection of a human lung showing a bronchus tube. c) A cut section of human bronchus tube from a transplanted lung after surgery. This section is incubated in Tyrode physiological solution for 18 h during which time it secrets 10 – 30 µl of mucus. d) Schematic of the epithelium including mucus secreting goblet cells, ciliated cells and mucus lining. Mucus contains mucin glycoproteins which form a porous network with pore sizes in the range of 100 – 5000 nm.[12] (e) Phase contrast microscopy image of mucus with injected micron-sized magnetic wires. The wires are marked with arrows. The mucus has a heterogeneous structure that is observable under phase contrast microscopy. The scale is 10 µm.

**2.7 - Comparison of magnetic rotational spectroscopy with cone-and-plate rheology**

The effects of viscosity and elasticity on the wire motion were also predicted from the constitutive equations associated with classical rheological models.[40] The validity of the MRS technique was confirmed by experiments performed on wormlike micellar solutions[51] and polysaccharide gels[52] as models of viscoelastic liquid and soft solid which are detailed in **Supporting Information S7.** Macrorheology experiments of the elastic and loss moduli $G'(\omega)$ and $G''(\omega)$ using commercial rheometers equipped with a cone-and-plate geometry were also performed for comparison. As a result, it was shown that MRS measures the linear rheology response parameters,[40] such as the static shear viscosity $\eta$ and the instantaneous and equilibrium elastic shear moduli, $G$ and $G_{eq}$ respectively.

# 3 - Results and Discussion

**3.1 - The different regimes of wire rotation in human respiratory mucus**

In this study, we investigated the rotation of a total of 98 magnetic wires dispersed in human mucus samples, 40 from the late *Ex Vivo* and 58 from the early *Ex Vivo* mucus. Each wire was studied at 4 to 5 frequency between $3\times10^{-3}$ and 30 rad s$^{-1}$. For these 98 wires, we found two generic rheological behaviors, namely that of viscoelastic liquids[51] and that of soft solids.[52] Interestingly, none of the wire studied showed rotation features of a purely viscous liquid.[58] This is because the sizes of the investigated wires were larger than the sizes of mucus microcavities. Magnetic wires indeed exhibit specific angular frequency dependencies whether they are embedded in one or the other of the materials mentioned above. For viscoelastic liquids, at low angular frequency $\omega$, the wires rotate synchronously with the field, and above a critical frequency $\omega_C$ they show asynchronous motion with back-and-forth oscillations. In that case, the average rotational velocity is different from zero at all values of $\omega$. For soft solids, a unique response is found as a function of the frequency, namely that of an asynchronous oscillation associated with $\Omega(\omega, t) = 0$. This specificity of soft solids by MRS comes from the fact that at the magnetic fields used, the stresses applied locally by the wires are lower than the yield stress and that we are dealing with in a purely deformational regime, and not with a flow regime.[52] Of note, each mucus sample studied exhibited both rheological behaviors, an outcome which will be explained later on the basis of a quantitative analysis of the $\theta(\omega, t)$-times series. Table I shows the number of wires entering each rheological class for late and early *Ex Vivo* mucus. We have found that samples taken from lung after surgery were contaminated with blood and possibly other lung fluids (**Supporting Information S1**). It was hypothesized that its rheological property may be impaired and do not reflect that of pristine mucus. For this reason, we focus first on the late *Ex Vivo* mucus which was collected from cultured bronchus tissues.

| Mucus type | Number of wires with viscoelastic liquid behavior | Number of wires with soft solid behavior | Total |
|---|---|---|---|
| Late *Ex Vivo* mucus | 24 | 16 | 40 |
| Early *Ex Vivo* mucus | 35 | 23 | 58 |

*Table I: Summary of the number of wires exhibiting viscoelastic liquid and soft solid rheological behaviors.*

### 3.2 - Late *Ex Vivo* human mucus with viscoelastic liquid behavior
*Viscosity measurements*

We will initially describe the features corresponding to the viscoelastic liquid behavior. Fig. 2a illustrates a 23 µm wire under a 10.3 mT rotating magnetic field and frequency equal to 0.0094 rad s$^{-1}$. One observes a heterogeneous background under phase contrast microscopy which is associated with a non-uniform material distribution. It is also noted that the sizes of the heterogeneous regions (~ 1 µm) are smaller than the length of the wire. The different images of the chronophotograph at time intervals of 100 s show that the wire rotates synchronously with the field. By increasing the frequency to 0.94 rad s$^{-1}$ (Fig. 2b), the wire exhibits a transition to asynchronous rotation whereby it performs back-and-forth oscillations: after an initial increase in the orientation angle in the direction of the field (yellow arrow), the wire undergoes a subsequent return in the opposite direction (red arrow) followed by a next rise in the orientation angle. We refer to the transition from the synchronous to the asynchronous regime by a frequency that we denote as critical frequency, $\omega_C$. For the wire in Fig. 2, $\omega_C$ is estimated to be 0.12 rad s$^{-1}$.

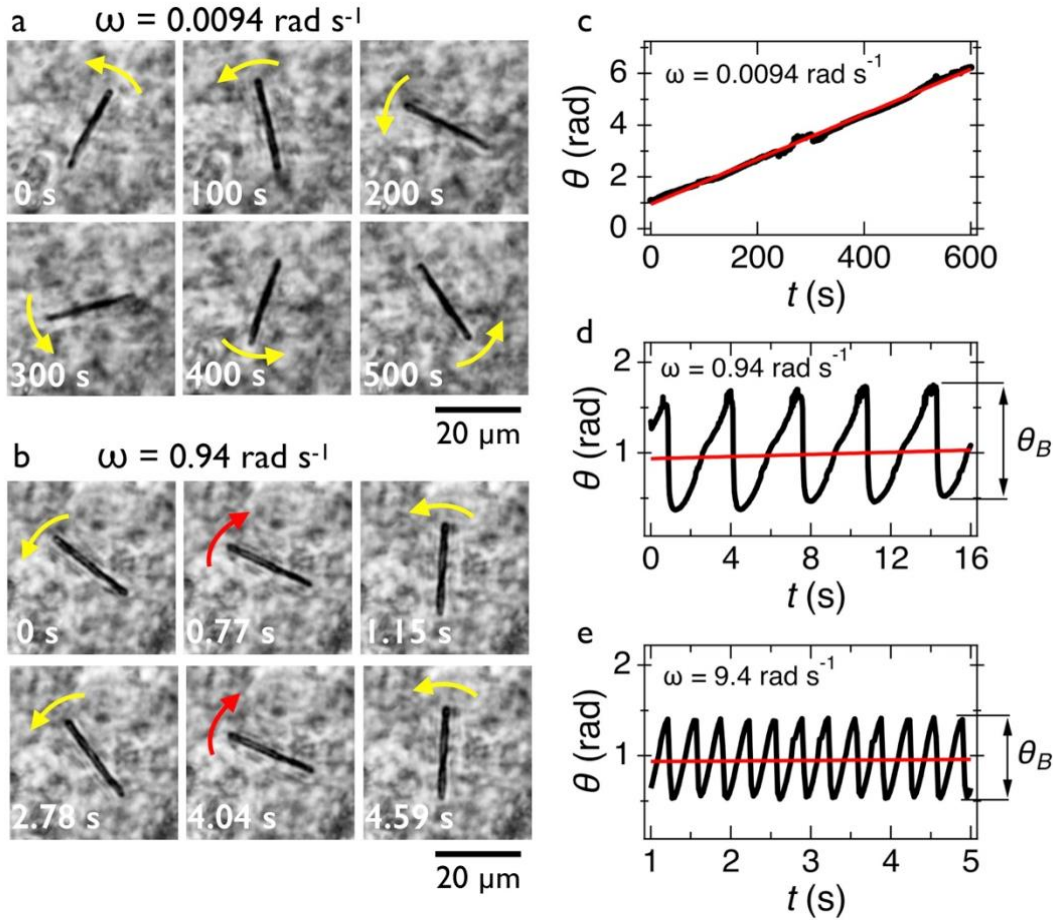

*Figure 2: **a-b**) Chronographs of a 23 µm wire undergoing rotation in late Ex Vivo mucus as a result of a 10.3 mT external magnetic field rotating with a frequency of 0.0094 rad s$^{-1}$ (a) and 0.94 rad s$^{-1}$ (b). **c-e**) Time series of orientation angle $\theta(t)$ in the synchronous (c) and asynchronous regimes (d and e). The average rotational velocity $\Omega(\omega)$ is depicted via straight red lines and the oscillation amplitude $\theta_B$ by arrows.*

In Fig. 2c, we show the $\theta(t)$-time series at $\omega = 0.0094$ rad s$^{-1}$ below the critical frequency. The orientation angle rises linearly with time and the slope of the temporal evolution is shown by a straight red line denoting $\Omega$. For $\omega < \omega_C$, we find $\Omega(\omega) = \omega$. In Figs. 2d and 2e, we show the $\theta(t)$-time series at $\omega = 0.94$ rad s$^{-1}$ and 9.4 rad s$^{-1}$, whereby $\Omega$ is now found to be 0.015 rad s$^{-1}$ and 0.006 rad s$^{-1}$, respectively (red lines). Above the critical frequency ($\omega > \omega_C$), $\Omega$ becomes smaller than $\omega$ and follows the relation[47,58] $\Omega(\omega) = \omega - \sqrt{\omega^2 - \omega_C^2}$. In the asynchronous regime, the oscillation amplitude $\theta_B$ as depicted by arrows is the angle by which the wire turns in the opposite direction to the field, here about 1.2 rad and 0.9 rad, respectively. The movies associated with Figs. 2a and 2b are provided in **Supporting Movies 1 & 2**.

Figs. 3a, 3b and 3c show the evolution of $\Omega(\omega)$ for wires with lengths 10, 24 and 33 µm, respectively. One observes that $\Omega(\omega)$ increases linearly with the actuation frequency in the synchronous regime ($\omega \leq \omega_C$) until it passes through a cusp-like maximum at the critical frequency and then falls rapidly ($\omega > \omega_C$). This behavior is used to interpret the critical frequency for each wire. Fig. 3d shows the normalized velocity $\Omega(\omega)/\omega_C$ as a function of normalized frequency $\omega/\omega_C$ for the 24 wires which share this behavior. One observes that the normalized wire responses agree reasonably with each other, especially considering the heterogeneity of mucus and its non-uniform material distribution. Increased dispersity at high frequencies is associated with a lower precision in the determination of $\Omega$ from wire tracking microscopy in this part of the spectrum.

The viscosity of mucus can be identified from the asymptotic behavior of $\omega_C$ as a function of the wire length. As shown previously,[51] $\omega_C$ is related with the static viscosity $\eta$ *via* the expression:

$$\omega_C = \frac{3}{8\mu_0} \frac{\Delta\chi}{\eta} \frac{B^2}{L^{*2}} \qquad (1)$$

where $L^* = L/[D\sqrt{g(L/D)}]$ is the reduced wire length, $g(x) = ln(x) - 0.662 + 0.917x - 0.050x^2$,[51] and $L$ and $D$ the wire length and diameter, $\mu_0$ the vacuum permeability, $B$ the magnetic field and $\Delta\chi$ the anisotropy of magnetic susceptibility. From the measurements of the diameter as a function of its length by optical microscopy, we calculated the reduced wire length, $L^* = L/[D\sqrt{g(L/D)}]$ and plotted it as a function of the actual length, leading to a scaling of the form $L^* = 1.663L^{0.637}$. Fig. 3e displays $\omega_C$ as a function of $L^*$ for the same wires shown in Fig. 3d. In this figure, each data point represents a single wire embedded in a different mucus environment. Using Eq. 1 we calculated the static viscosity of the individual wires and obtained a mean value equal to 100 ± 40 Pa s (± error of mean, 6 – 280 Pa s are the limits of the shaded area in Fig. 3e). The viscosity we measure is much higher than those reported by Schuster et al.[10] using particle tracking (0.05 – 3.4 Pa s for probe sizes between 100 and 500 nm) and also slightly higher than that of sputum (67 ± 34 Pa s) obtained by macrorheology (see Section 3.5 *Comparison with literature data*). This outcome signifies the success of the MRS technique in measuring the viscosity of a heterogeneous sample from the local rotation-actuation response of magnetic wires.

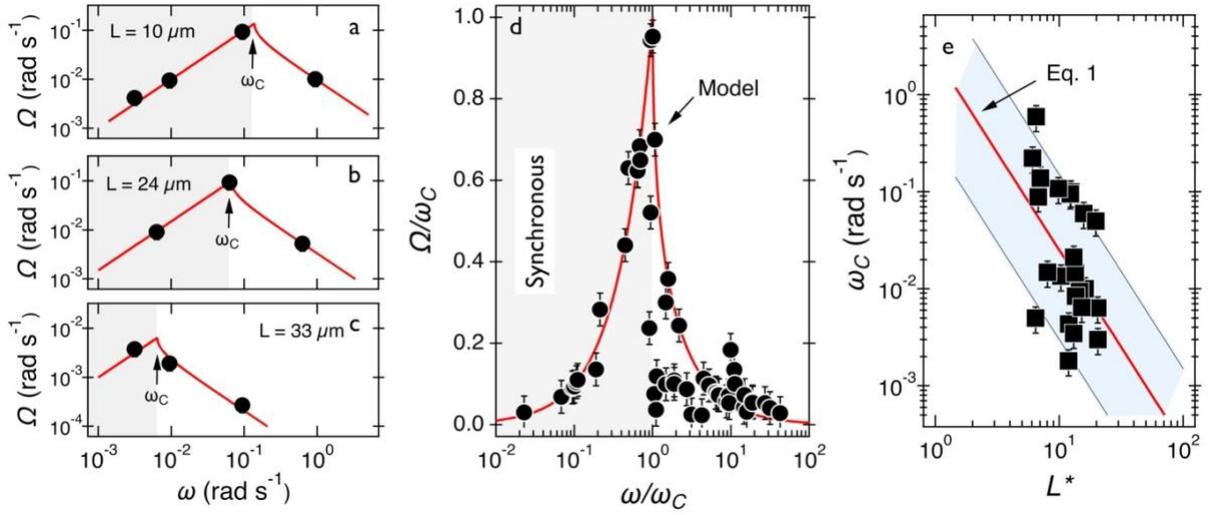

***Figure 3**: **a-c)** Average rotational velocity of several wires with length between 10 and 33 µm. Solid fit lines are from the expressions given in the text. **d)** Average rotational velocity of wires normalized by the critical frequency $\Omega(\omega)/\omega_C$ plotted versus normalized frequency $\omega/\omega_C$. The continuous line comes from the model developed for fluids of finite viscosity[51,58] **e)** Variation of critical frequency $\omega_C$ as a function of reduced wire length $L^*$. The thick solid line shows the $1/L^{*2}$-dependence corresponding to a mean viscosity $\eta$= 100 ±40 Pa s.*

*Elastic modulus measurements*

Figs. 4a, 4b and 4c show $\theta_B$ as a function of $\omega$ for wires with lengths 12, 27 and 52 µm, respectively. As a general behavior, $\theta_B$ decreases with frequency and we denote the value at the highest actuation frequency by $\theta_0$. Fig. 4d displays $\theta_B(\omega)$ as a function of the normalized frequency $\omega/\omega_C$. It is found that the data superimpose over 4 decades in frequency and follow a scaling law with exponent -0.10. On the same figure we also show the solution of the constitutive equation for Newton fluids, which predicts a fast fall of $\theta_B$ following the relation $\theta_B(\omega/\omega_C) \sim (\omega/\omega_C)^{-1}$ for $\omega \gg \omega_C$. The variation of the oscillation amplitude as observed with wires in this regime is a signature of a viscoelastic liquid.[44,51] The elastic modulus can be interpreted from the asymptotic behavior of $\theta_0$ as a function of the wire length. As shown previously,[51] this angle is related with the elastic modulus $G$ *via* the expression:

$$\theta_0 = \frac{3}{4\mu_0} \frac{\Delta\chi}{G} \frac{B^2}{L^{*2}} \qquad (2)$$

Fig. 4e displays this angle for the 24 individual wires of this class as a function of the reduced wire length and also depicts the $1/L^{*2}$-dependence of this angle. The shaded area in this figure shows the extent of variation of the prefactor in Eq. 2. Using Eq. 2, we calculated the elastic modulus and obtained a mean of $G = 2.5 \pm 0.5$ Pa (0.5 – 13 Pa are the limits of the shaded area). The variation in elasticity values similar to the viscosity is due to mucus heterogeneity at the scale of the wire length, *i.e.* in the range 5 – 80 µm.

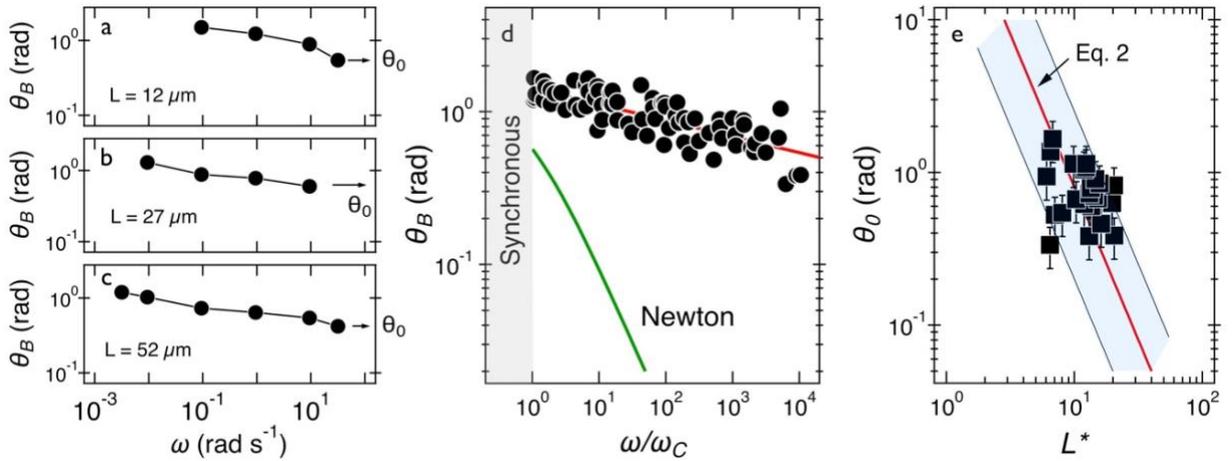

*Figure 4: a-c)* Oscillation amplitude of several wires with lengths 12 – 52 µm. Arrows point at the oscillation amplitude at the highest frequency, denoted by $\theta_0$. *d)* Oscillation amplitude $\theta_B$ as a function of normalized frequency $\omega/\omega_C$ and comparison with that of a Newton fluid. *e)* Variation of $\theta_0$ as a function of reduced wire length $L^*$. Solid line shows the $1/L^{*2}$-dependence of Eq. 2, leading to $G = 2.5 \pm 0.5$ Pa. In this figure, each data point represents a different wire in mucus.

### 3.3 - Late *Ex Vivo* human mucus exhibiting soft solid characteristics

Out of 40 wires investigated in late *Ex Vivo* mucus, 16 wires with length between 8 and 75 µm showed a soft solid behavior characterized by an average rotational velocity close to zero, or precisely $\Omega(\omega) < \Omega_{\text{Res}} = 10^{-4}$ rad s$^{-1}$, at all values of $\omega$. This limitation is obtained from the resolution of the optical microscope and that of the wire orientation obtained from the tracking ImageJ plugin.[49] During observation periods which can be longer than 1 h the wire remained stationary and only performed an oscillatory motion. Figs. 5a and 5b show examples of such behavior for a wire with length 23 µm under actuation fields with $\omega = 0.0031$ rad s$^{-1}$ and 0.0094 rad s$^{-1}$ respectively. These observations are generally indications that a viscoelastic (soft) solid material surrounds the wire. From this type of response, a critical frequency cannot be assigned. It is possible however to obtain $\theta_B(\omega)$ and calculate $G$ from Eq. 2. Fig. 5c displays $\theta_B(\omega)$ for wires with lengths between 14 µm and 75 µm and Fig. 5d shows the $\theta_0(L^*)$-dependence that corroborates the scaling of Eq. 2.[52] We calculated $G$ for individual wires and obtained a mean of $G = 2.6 \pm 0.3$ Pa. A similar calculation was made for the low frequency region, from which the equilibrium modulus can be derived *via* the relation:[44,52]

$$\lim_{\omega \to 0} \theta_B(\omega) = \theta_{eq} = \frac{3}{4\mu_0} \frac{\Delta\chi}{G_{eq}} \frac{B^2}{L^{*2}} \tag{3}$$

It is found that $G_{eq} = 1.7 \pm 0.3$ Pa, *i.e.* a value close to that of $G$. The movies associated with the wire rotations in Figs. 5a and 5b are shown in **Supporting Movies 3 & 4**.

We thereby find that the elastic moduli measured at regions with local viscoelastic liquid or soft solid behavior are in good agreement. This agreement shows that the elastic networks of these mucus regions with these unlike behaviors are similar. We therefore assume that these differences come from the relaxation times. For the viscoelastic fluid case, the relaxation time of mucus can be estimated from the values of the viscosity and the elastic modulus, $\tau = \eta/G$ for each wire, leading to a mean relaxation time equal to $\tau = 47 \pm 12$ s (min = 1 s, max = 290 s). For the soft solid case, we can assign an upper limit for $\omega_C$, of the order of the minimum angular frequency used *i.e.* $10^{-3}$ rad s$^{-1}$. This outcome indicates that the current MRS technique is limited to measurements of relaxation times less than 300 s. To probe such long relaxation times, a steady rotation of wires could be achieved if the wire length would be reduced but this has the disadvantage of approaching the size of microcavities, or if the angular frequency would be decrease below $10^{-3}$ rad s$^{-1}$.

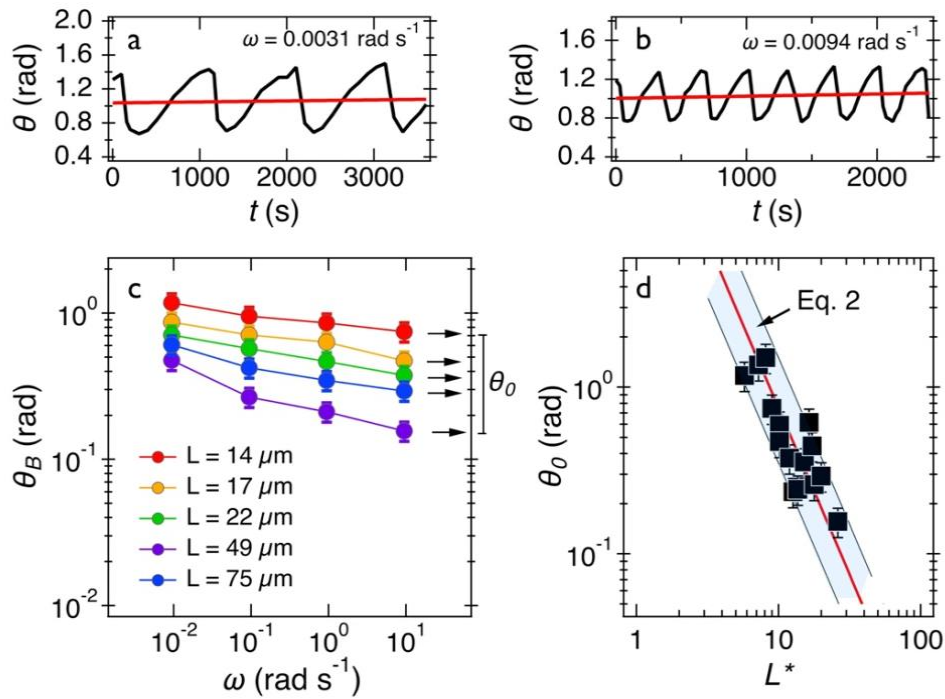

*Figure 5*: *a, b*) *Time dependence of orientation angle $\theta(t)$ for a wire experiencing the second generic behavior at actuation frequencies 0.0031 rad s$^{-1}$ and 0.0094 rad s$^{-1}$ respectively. c) Oscillation amplitude of wires with lengths 14 – 75 µm. Arrows point at the amplitude at the highest frequency. d) Variation of oscillation amplitude $\theta_0$ as a function of the reduced wire length. Solid line shows the $1/L^{*2}$-dependence of Eq. 2, leading to G = 2.6 ± 0.3 Pa. In this figure, each data point represents a different wire dispersed in mucus.*

### 3.4 – Early *Ex Vivo* human mucus collected after surgery

We also investigated mucus that was directly collected from the bronchus tube after surgery which we called the early *Ex Vivo* mucus. Similarly, we find two generic behaviors from the wire rotations. Out of the 58 wires investigated in 8 samples, 35 presented the response of a viscoelastic liquid and 23 showed the response of a soft solid. Note that the proportion between the two types of responses was similar to that of the late *Ex Vivo* mucus, *i.e.* around 60% (Table I). In Fig. 6 we compare the behavior of early and late *Ex Vivo* mucus regarding the viscoelastic liquid behavior found for these 35 wires. The data received for the average rotational velocity $\Omega(\omega)$, the oscillation amplitude $\theta_B(\omega)$, the critical frequency $\omega_C$ and for the measured oscillation angle $\theta_0$ extrapolated to $\omega \to \infty$ are plotted in Figs. 6a, 6b, 6c and 6d, respectively. Fig. 6a shows the average rotational velocity $\Omega(\omega)$ normalized with the critical frequency $\omega_C$ as a function of the reduced actuation frequency $\omega/\omega_C$. As for late *Ex Vivo* mucus samples, the data are well accounted for by the viscoelastic model prediction (continuous curve).[47,58]

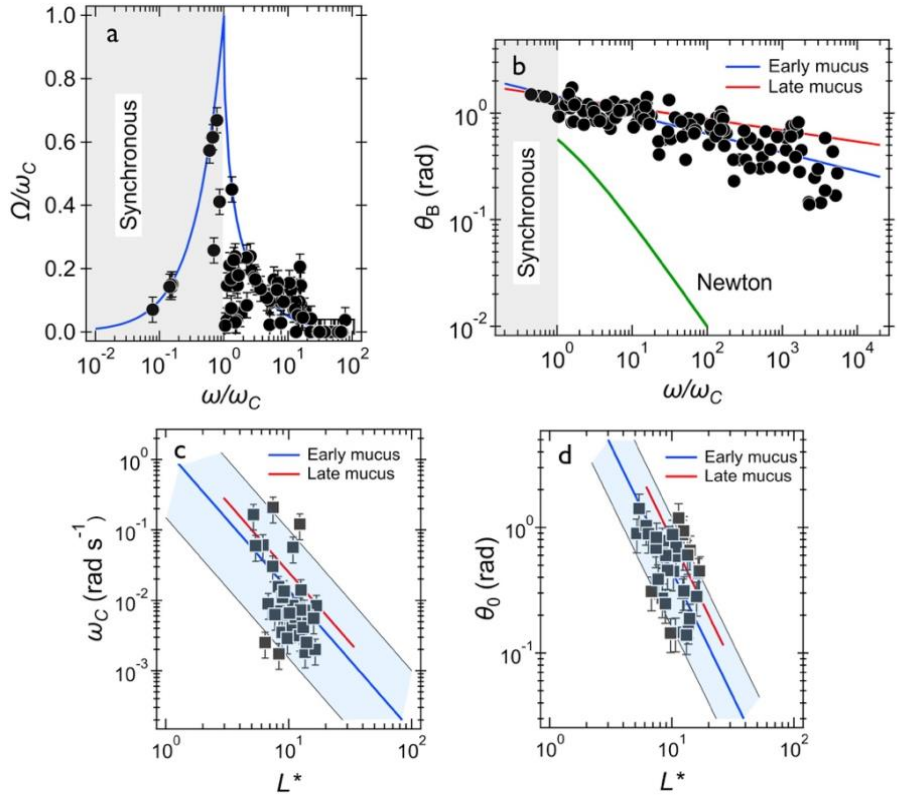

*Figure 6*: Wire microrheology results obtained for early Ex Vivo human mucus. *a)* Average rotational velocity of wires normalized by their individual critical frequency $\Omega(\omega)/\omega_C$ plotted versus normalized frequency $\omega/\omega_C$. The response behaviors of the wires (35 wires) agree with the model developed for fluids of finite viscosity (continuous line in blue)[51,58]. *b)* Oscillation amplitude $\theta_B(\omega)$ as a function of $\omega/\omega_C$. Also depicted are the prediction for a Newton liquid (continuous curve in green) and best fit calculations obtained for the late Ex Vivo mucus in reference with the data of Fig. 4d (straight line in red). *c)* Variation of critical frequency $\omega_C$ as a function of reduced wire length $L^*$. The thick solid line in blue shows the $1/L^{*2}$-dependence corresponding to a mean viscosity $\eta = 180 \pm 30$ Pa s. *d)* Variation of $\theta_0$ as a function of reduced wire length $L^*$. The thick solid line in blue shows the $1/L^{*2}$-dependence of Eq. 2, leading to $G = 4.4 \pm 0.5$ Pa. In Fig. 6c and 6d, the straight lines in red result from best fit calculations obtained for late Ex Vivo mucus (data from Figs. 3e and 4e, respectively).

Fig. 6b displays $\theta_B(\omega)$ as a function of the normalized frequency $\omega/\omega_C$. It is found that the data superimpose over 4 decades in frequency and follow a scaling law of the form $\theta_B(\omega) \sim (\omega/\omega_C)^{-0.18}$. Also shown on the figure is the outcome for the late *Ex Vivo* mucus, represented by straight line in red with an exponent -0.10. The decrease in $\theta_B(\omega)$ with the frequency is slightly more pronounced for the early *Ex Vivo* mucus (exponent -0.18 *versus* - 0.10), the two set of data being however close to each other. Fig. 6c displays $\omega_C$ as a function of $L^*$ for the same 35 wires. It is found that the scattering of the data points is important and the data do not exactly follow the $1/L^{*2}$-prediction. As for the late *Ex Vivo* mucus, this outcome is ascribed to mucus heterogeneities that are observed on the viscosity. Using Eq. 1 we calculated the static viscosity of each individual wire and obtained a mean value equal to 180 ± 30 Pa s (straight line in blue), again in good agreement with the late *Ex Vivo* data (straight line in red). The data for the oscillation angles $\theta_0$ *versus* $L^*$ are depicted in Fig. 6d, and they are basically alike for the two mucus samples. Using Eq. 2, we calculated the elastic modulus and obtained a mean of 4.4 ± 0.5 Pa. We assign a relaxation time to the first generic behavior equal to $\tau$ = 49 ± 6 s and an upper limit for the relaxation time in the second generic behavior estimated to be > 200 s. In the second generic behavior (*i.e.* that of the soft solid) we calculated an elastic modulus equal to $G_{eq}$ = 2.0 ± 0.2 Pa (**Supporting Information S8**). In conclusion to this part, we find that the viscoelastic properties of early *Ex Vivo* mucus agree reasonably well with those calculated in late *Ex Vivo* mucus, indicating that the contamination identified in samples taken after surgery have little to no effect on the rheology. In addition, the late *Ex Vivo* mucus is similar in production to cell culture methods certifying the applicability of our protocol in investigations related to mucus flow in airway surfaces.

| Mucus type | Viscoelastic liquid behavior | | | Soft solid behavior |
|---|---|---|---|---|
| | $\eta$ (Pa s) | $G$ (Pa) | $\tau$ (s) | $G_{eq}$ (Pa) |
| Late *Ex Vivo* mucus | 100 ± 40 | 2.5 ± 0.5 | 47 ± 12 | 2.6 ± 0.3 |
| Early *Ex Vivo* mucus | 180 ± 30 | 4.4 ± 0.5 | 49 ± 6 | 2.0 ± 0.2 |

*Table II*: *Combination of the rheological parameters obtained for late and early Ex Vivo mucus using the magnetic rotational spectrometry technique. Here, $\eta$, $G$ and $\tau$ denote the static shear viscosity, the elastic modulus and the relaxation time, respectively.*

### 3.5 - Comparison with literature data

Following this study, it was important to compare our *Ex Vivo* mucus microrheology data to that of the literature. To this end, macro and microrheological data from 23 sputum and mucus surveys were evaluated.[9-11,13,14,16,24,26-39] The nature, origin and references of the samples selected for this analysis are listed in **Supporting Information S9**. From these reports, the elastic modulus $G'(\omega)$, the loss modulus $G''(\omega)$ and the complex modulus $G^*(\omega) = \sqrt{G'(\omega)^2 + G(\omega)''^2}/\omega$ were compiled for cystic fibrosis sputum and healthy mucus. Whenever this was possible, we have taken into account the fact that the above elastic moduli depend on the measuring frequency. To facilitate comparison, data obtained at different frequencies were extrapolated to $\omega$ = 1 rad s$^{-1}$ using the scaling laws found by Schuster e*t al.* [10] on healthy human mucus, *i.e.* $G'(\omega) \sim \omega^{0.15}$ and $G''(\omega) \sim \omega^{0.15}$, and identical expressions were assumed for sputum and mucus. For cystic fibrosis sputum (for which the number of samples were the most numerous), averaged moduli $G'$, $G''$ and $G^*$ were obtained from $n = 8$, 6 and 9 independent studies, respectively. It was found that at 1 rad s$^{-1}$, $G' = 4.4 \pm 1.9$ Pa, $G'' = 3.0 \pm 2.0$ Pa and $G^* = 11.1 \pm 4.0$ Pa. For mucus samples, the following results were obtained: $G' = 7.3 \pm 2.9$ Pa, $G'' = 2.1 \pm 0.7$ Pa and $G^* = 10.7 \pm 2.6$ Pa from $n = 3$, 3 and 5 independent studies, respectively. Finally, the shear viscosity obtained at low shear rate (< 0.1 s$^{-1}$) was also evaluated and found to be $67 \pm 34$ Pa s for COPD and cystic fibrosis sputum ($n = 5$). Pertaining to human mucus, there is not enough stationary viscosity data available in the literature to draw a clear conclusion at this point. Figs. 7a and 7b show the moduli histograms ($G'$ in dark blue, $G''$ in green and $G^*$ in orange) for the cystic fibrosis sputum and healthy mucus, whereas Figs. 7c compares the shear viscosity found in this work with that of sputum. Given the large uncertainties from the literature data, our results (in red in Figs. 7b and 7c) are in overall agreement with those reported in earlier studies, demonstrating that the MRS method is reliable and sensitive. Furthermore, our measurements reveal the existence of a wide varying relaxation times in human respiratory mucus beyond those that could be related with the microcavities. The biological significance of this variation (e.g. in terms of an effective mucociliary clearance) is not clear at the moment. We however suspect that it originates from different local secretions of mucus and electrolytes by the respiratory epithelial cells.

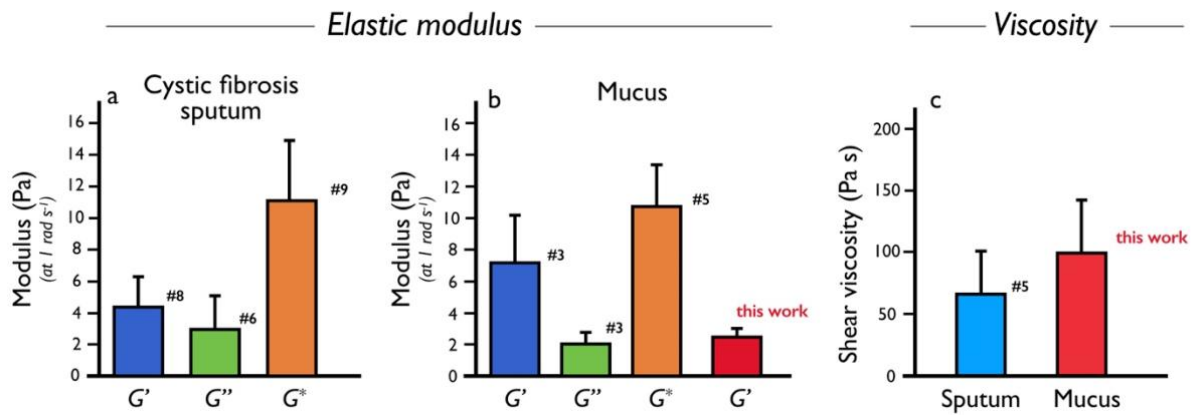

*Figure 7: a) Elastic ($G'$, dark blue), loss ($G''$, green) and complex ($G^*$, orange) moduli data collected from the literature for human cystic fibrosis sputum. The values of the moduli shown are for an angular frequency of 1 rad s-1. The number close to the bars denote the number of references used to calculate the average values and standard errors. The list of references used to establish these statistics, as well as the related values of the moduli and viscosities are presented in the Supporting Information S9. b) Same as in a) for human mucus. The bar in red in from this work (Table II). c) Same as in a) for shear viscosity of human sputum and mucus (this work).*

## 4 - Conclusion

In this work we propose a new active microrheology technique to determine the rheological characteristics of human respiratory mucus. The probes used are remotely actuated magnetic wires of lengths between 5 and 80 µm, *i.e.* much greater than the size of the microcavities (100 nm - 5 µm) commonly observed in mucus samples.[12] We thereby ensure that the viscosity and elasticity values obtained are averaged over length scales large enough to be compared with the bulk values. One result that emerges from this work is that human respiratory mucus is a viscoelastic material, an outcome that corroborates earlier literature reports.[5] The ability to perform MRS experiments at angular frequency as low as $10^{-3}$ rad s$^{-1}$ makes it possible to highlight the viscoelastic liquid character of the mucus, and also to estimate for the first time the relaxation times of the mucin network. In human respiratory mucus, we have found two types of rheological behavior, namely that of a viscoelastic liquid and that of a soft solid. Quite remarkable, these two behaviors are observed on the same samples, typically in a ratio 60%-40%. While a similar elastic modulus was obtained in viscoelastic liquid and soft solid regimes, of the order of 2 - 5 Pa, the relaxation times (the ratio between the viscosity and the modulus) are different, one being measurable at about 40-50 s, and the other being about an order of magnitude longer (> 300 s). It is concluded that the variability in viscosity that we obtain is related with the spatial distribution of relaxation times in mucus at a scale comprised between 10 and 100 µm (*i.e.* the size of the wires), an outcome that was not previously disclosed and which could be relevant for mucus thinning treatments. We also showed that mucus collection from tissue culture, leading to the late *Ex Vivo* mucus in this work can be successfully used for the investigation of mucus rheology. This suggests that the tissue culture method, but also possibly culture of human lung epithelial cells may be appropriate for investigations related to drug delivery with similar rheological properties to *In Vivo* mucus. This work finally shows that beyond the established structural variations due to the microcavities, mucus exhibits secondary inhomogeneities associated with the relaxation times of the mucin network that may be important for its flow properties.

# Acknowledgments

ANR (Agence Nationale de la Recherche) and CGI (Commissariat à l'Investissement d'Avenir) are gratefully acknowledged for their financial support of this work through Labex SEAM (Science and Engineering for Advanced Materials and devices) ANR 11 LABX 086, ANR 11 IDEX 05 02. We acknowledge the ImagoSeine facility (Jacques Monod Institute, Paris, France), and the France BioImaging infrastructure supported by the French National Research Agency (ANR-10-INSB-04, « Investments for the future »). This research was supported in part by the Agence Nationale de la Recherche under the contract ANR-13-BS08-0015 (PANORAMA), ANR-12-CHEX-0011 (PULMONANO), ANR-15-CE18-0024-01 (ICONS), ANR-17-CE09-0017 (AlveolusMimics) and by Solvay.

# Supporting Information

S1: Human mucus samples used in this study – S2: Origin of the human mucus samples – S3: Synthesis of magnetic wires – S4: Characterization of magnetic wires – S5: Effect of pH on the magnetic wire stability – S6: Magnetic rotational spectroscopy: methods: S7 – Magnetic rotational spectroscopy *versus* macrorheology: a comparative study – S8: Analysis of early *Ex Vivo* mucus – S9: Review of previous sputum and mucus rheology

# TOC and graphical abstract

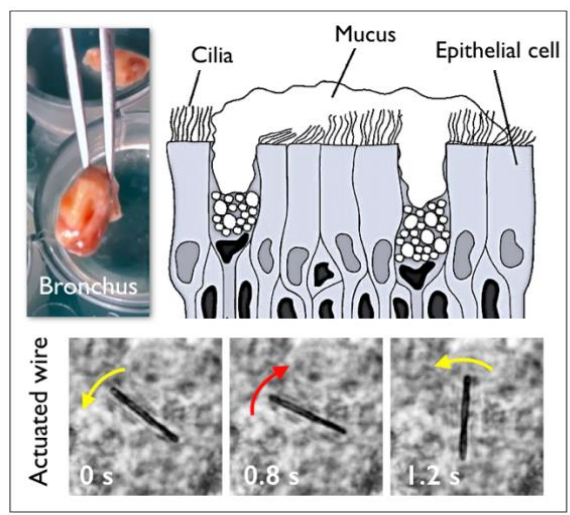

Magnetically actuated micron-sized wires are used to address the question of the mechanical properties of mucus gels collected from human bronchus tubes following surgery. Our work shows that mucus has the property of a high viscosity gel characterized by large spatial viscoelastic heterogeneities.